\documentclass[12pt,a4paper,twoside]{article}
\usepackage{epsfig}
\usepackage{baltlat1}
\begin{document}
\ \
\vspace{0.5mm}

\setcounter{page}{1}
\vspace{8mm}

% laur LAUR 03-8609
\titlehead{Baltic Astronomy, vol.12, XXX--XXX, 2003.}

\titleb{SWIFT'S ABILITY TO DETECT GAMMA-RAY BURSTS}

\begin{authorl}
\authorb{E.~E.~Fenimore}{1}
\authorb{K.~McLean}{1,2}
\authorb{D.~Palmer}{1}
\authorb{S.~Barthelmy}{3}
\authorb{N.~Gehrels}{3}
\authorb{H.~Krimm}{3}
\authorb{C.~Markwardt}{3}
\authorb{A.~Parsons}{3}
\authorb{M.~Stephens}{3}
\authorb{and J.~Tueller}{3}
\end{authorl}

\begin{addressl}
\addressb{1}{Los Alamos National Laboratory,
MS B244, New Mexico 87545, USA}

\addressb{1}{University of Texas at Dallas, Texas, 75080 USA}

\addressb{3}{Goddard Space Flight Center, Greenbelt Maryland USA}
\end{addressl}

\submitb{Received November 15, 2003}

\begin{abstract}

The Swift satellite will be a self-contained observatory that will bring
new capabilities to the observing of the early afterglow emission of
Gamma-ray Bursts. Swift is completely autonomous and will do all of the 
observations without help from the ground. There are three instruments on 
Swift.  A large (5200 sq cm) coded aperture imager will locate the bursts
within about 15 seconds.  The satellite will be able to slew to point at
the location within a minute or two. There are two narrow field of view
instruments: an optical telescope and an x-ray telescope. Thus, Swift will
provide simultaneous gamma-ray, x-ray, and optical observations of
Gamma-ray bursts soon after the burst.

A key to the success of Swift will be its ability to detect and locate a large
number of gamma-ray bursts quick enough that the narrow field of view instruments
can follow up.  The results of simulations show that Swift will be able to detect
about 300 bursts a year and locate about 150. The number that Swift will be able to
slew to depends on constraints built into the satellite bus. Preliminary results
indicate that we might be able to slew to 100 bursts per year, but that is heavily 
dependent on satellite operations.
\end{abstract}


\resthead{Swift: An Autonomous Satellite}{E.~E.~Fenimore et al.}

%{Institution}{Author(s)}

%\def\ninepoint{\def\rm{\fam0\ninerm} \textfont0=\ninerm}

\sectionb{1}{INTRODUCTION}

 The Swift satellite (Gehrels et al, 2003)  will be the next mission
dedicated to the study of gamma-ray bursts. Launch is scheduled for
mid-2004. The key concept behind Swift is that it will ``catch gamma-ray
bursts on the fly'', much like the bird known as the swift catches insects
in flight.  Without any involvement from the ground team, the on-board
software will detect that a gamma-ray burst has started, locate it, and
slew the satellite so the x-ray and UV/optical telescopes can observe the
burst and its afterglow.

\begin{figure}
\vskip2mm
% apparently the height/width is set first before angle in epsfig
\centerline{\epsfig{figure=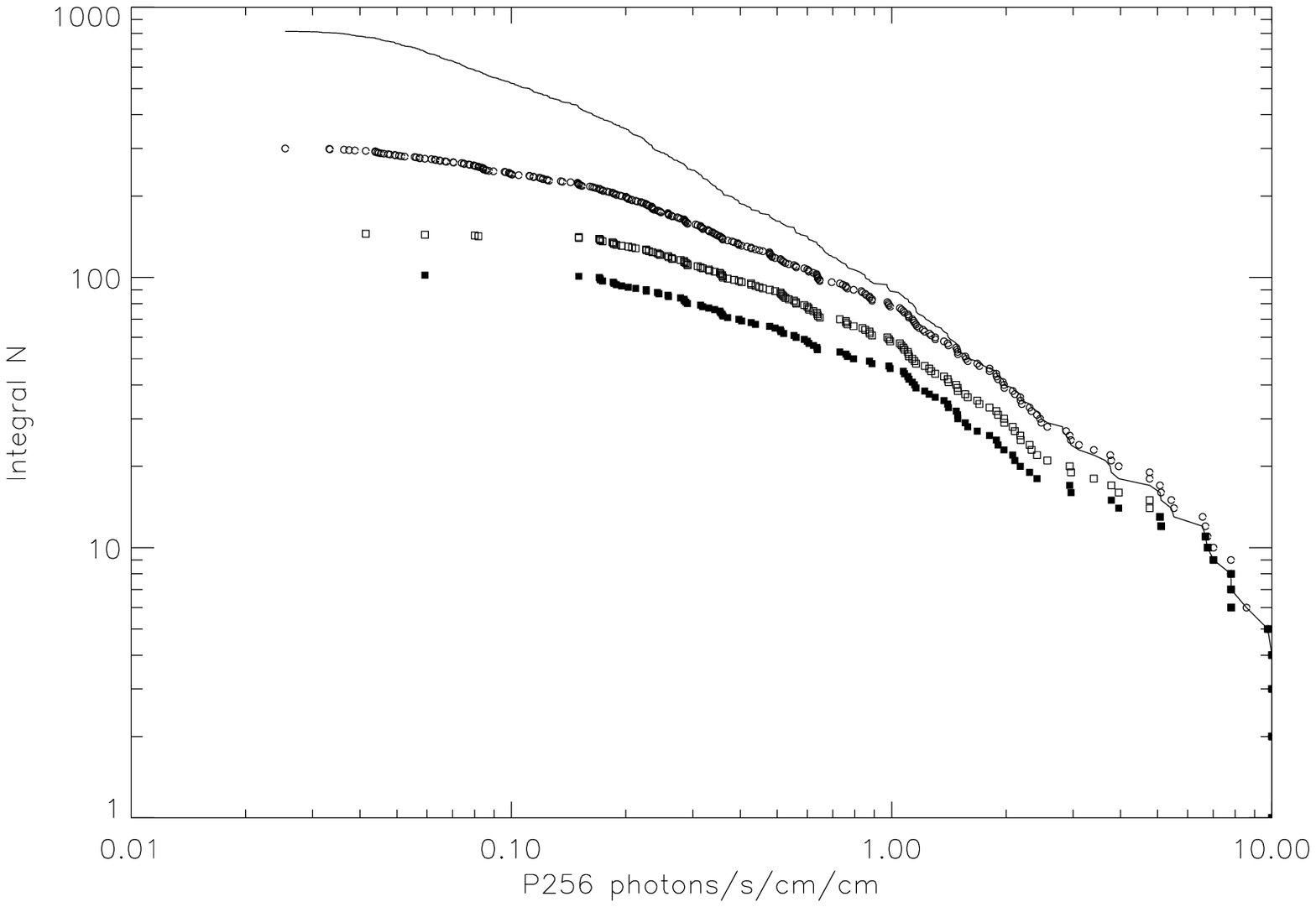,width=90truemm,angle=0,clip=}}
\captionb{1}{Log N Log P distribution from the 1000 burst simulation. The solid line  is 
the distribution of simulated bursts. The open circles are bursts that caused triggers while the 
open squares are bursts that were located. The solid squares are bursts that were slewed to. 
This simulation represented about a year but is heavily dependent on the assumed satellite 
operations. See Fig. 4 for an improved simulation.} \vskip2mm \end{figure}

The key to catching gamma-ray bursts while they are occurring is to rapidly locate the
gamma-ray burst. Swift's ``Burst Alert Telescope'' (BAT) is a large 15 to 150 KeV coded
aperture locator (see Barthelmy et al, 2003).  BAT has a 5200 cm$^2$ focal plane consisting of
32,678 CZT detectors (each 4 mm by 4 mm by 2 mm).  The aperture is made up of 54,000 lead
blocks, each 5 mm by 5 mm by 1 mm. The separation of the focal plane and the aperture is 100 cm which
translates into a point spread function of about 17 arcmin.

The BAT on-board flight software (Palmer 2004) ``triggers'' to detect
that a burst is happening. The trigger actually serves two purposes.  
First, detect that a burst is occurring based on the light curve in the
focal plane and, second (and more importantly), identify the best times to
use for the imaging. The signal to noise in the coded aperture image are
usually smaller by a factor of ~0.7 than the signal to noise in the light
curve from the focal plane. There will be statistically significant
triggers that do not have statistically significant images. To make the
best of this situation, it is important to find the period of time with
the absolutely highest signal to noise.

\begin{figure}
\vskip2mm 
% apparently the height/width is set first before angle in epsfig
\centerline{\epsfig{figure=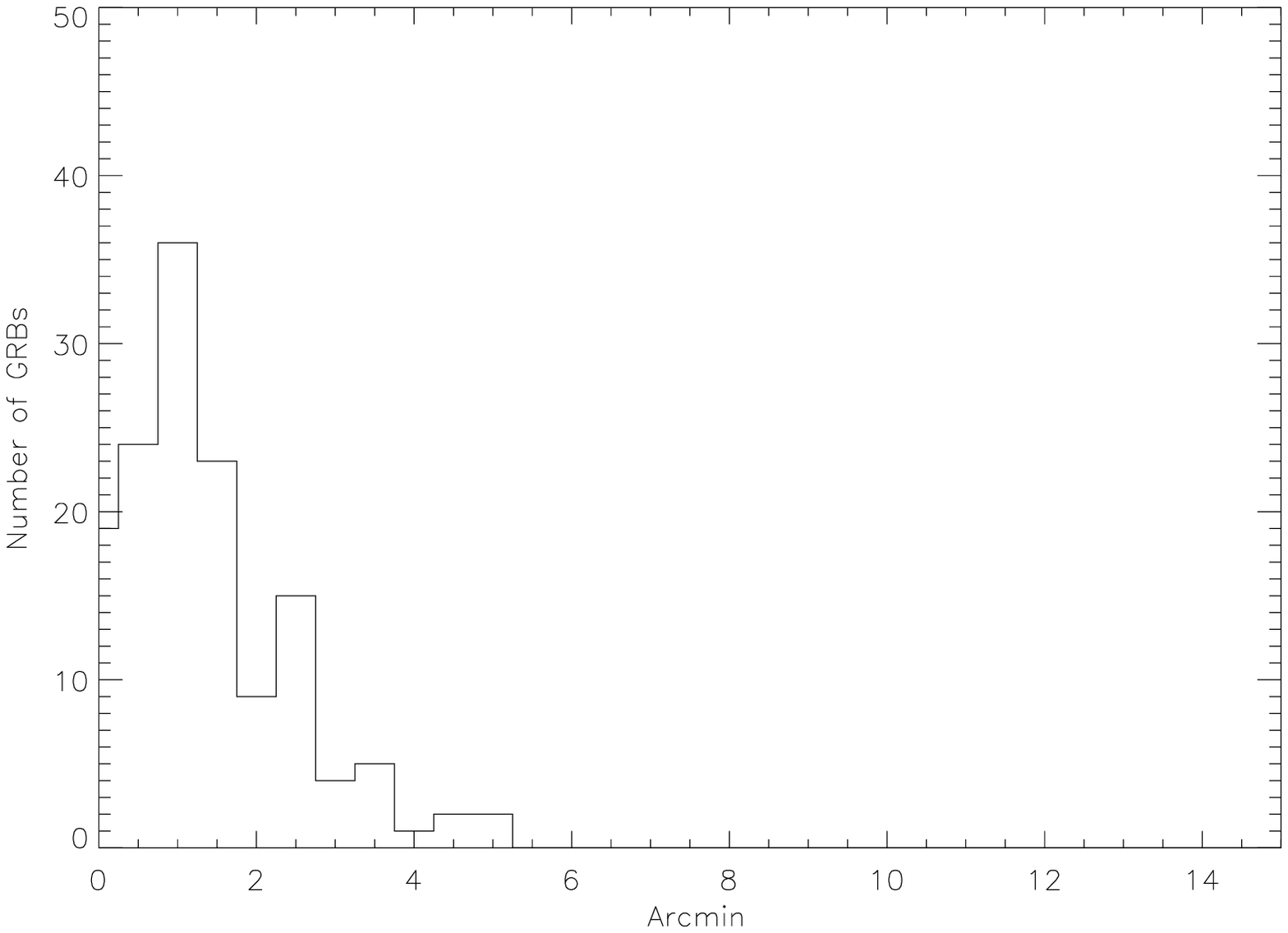,width=90truemm,angle=0,clip=}}
\captionb{2}{Accuracy of the on-board locating algorithm based on
simulating 1000 bursts. Using deconvolution coupled to a cleaning step
plus back project, the completely autonomous imaging software locates
about half the events to within 1.25 arcmin and 90\% with 3 arcmin.  The
FWHM of the point spread function is about 17 arcmin. See Fig. 5 for an improved simulation.} 
\vskip2mm
\end{figure}

Thus, to make the best images for locating, it is important to search
``all'' time scales and energy ranges.  We have several different
triggering systems (Fenimore et al 2000) to cover a wide parameter space.
One system is specialized for short bursts (4 ms to 64 msec).  This
``short'' trigger system covers 180 different combinations of energy
ranges, time scales, and focal plane regions. A ``long'' trigger systems
evaluates about 400 criteria covering from 64 msec to about ~8 seconds.
The long criteria can use two background samples to fit a linear function
to the background to remove trends. Although effective at eliminating
false triggers, in many cases the algorithm must wait until well after the
burst in order to measure the second background.
To speed up the locating, we have another 200 criteria that are activated
whenever there is a trigger. These additional criteria only use a single
background and do not have to wait until after the burst to find a large
signal. They are so effective in speeding up the location that we call
these ``afterburners''.

Every 64 sec, 320 sec, and for each pointed observation, we make an on-board image of the 
field of view and search for new sources. These ``image'' triggers will detect slow rising 
gamma-ray bursts and other 
transients.

\sectionb{2}{1000 BURST SIMULATIONS}

To test the flight software and the estimate the performance of Swift, we
run a ``1000 burst'' simulation. We randomly pick locations for 1000
gamma-ray bursts (within 63 degrees of the centerline of BAT), randomly
pick a time history (from a collection of bright BATSE events), randomly
pick a spectrum (from the distribution of Band parameters seen by Ginga),
and inject them into a simulated data stream from the focal plane.  We
also inject the x-ray diffuse background, steady sources, and orbital
variations in the particle background. The typical quiet orbit has count
rates that vary from 12 KHz to 32kHz while the South Atlantic Anomaly (SAA) 
will have count rates
up to 4 MHz. The simulated data stream from the focal plane is then run
through a ground copy of the flight computers. Other computers simulate
how and when the spacecraft will slew.

%  \subsectionb{1.1}{What we cannot obtain\hfil\break practically}

 The 1000 burst simulation covers about a year of operation. The peak
fluxes for each gamma-ray burst (GRB) is selected from the BATSE Log N Log
P distribution, analytically extended to lower fluxes. The upper Log N Log
P curve in Figure 1 is the distribution of randomly selected peak fluxes
in photons s$^{-1}$cm$^{-2}$ measured over 256 ms in the energy range 50
to 300 KeV (i.e., BATSE's P256). Although 1000 bursts were simulated
within 63 degrees of the center of the BAT field of view (FOV), 188 were
behind the earth and not seen. There were 400 triggers,of which about 100
were false triggers. Our strategy is to have low thresholds and to allow a fair
number of false triggers. The thresholds were set to allow a few false
rate triggers per orbit (McLean 2004).  During the simulation, as would be
true in flight, it is hard to tell if a trigger is a false rate trigger on
noise or a true trigger on a weak GRB that we could not locate. For that
reason, the number of bursts that we will detect, but not locate, is
uncertain but probably lies between 250 and 350.  Virtually none of the
false triggers form an image so we do not slew to them.

There were 148 GRB locations found. More extensive ground analysis based
on real time images through TDRSS could allow for a few more locations
within about 5 minutes.  The spacecraft constraints (earth limb, sun,
moon, etc) allowed us to slew to 103 of them within the first ~1000 sec.
We only simulated the first 1000 sec after the location so we do not know how
many of the 148 would have eventually been observed. How many we will
actually slew to in flight will be heavily dependent on satellite
operations.

Figure 2 shows the distribution of the difference between where the flight
software located the burst and their true locations. The on-board software
does very well: 90\% of the events were found within 3 arcmin and 50\% of
the events are found within 1.25 arcmin (radius). Given that the full
width half maximum (FWHM) of the point spread function is 17 arcmin, this
is a remarkable achievement for completely autonomous imaging software on
a 25 MHz processor. 

The software requires about 12 sec to produce and analyze an image. After
triggering, it continues to form images and search for new objects until
it finds a burst.  In about half the cases, it finds the burst within the
first image. Most are found within two images.  Once informed of a new
location, the spacecraft takes 20 to 100 seconds to slew and settle. In
some cases, the software will locate the burst during the early phase of a
long GRB and be able to slew to the burst while it is still happening. For
10 out of the 103 bursts that we slewed to, we arrived within T90 seconds
of the peak of the burst. Figure 3 gives how long after the peak of the
burst was required before the x-ray and UV/optical telescopes were able to
start their observations of the GRB. Some arrive with 30 seconds of the
peak, most took about 100 seconds.

\begin{figure}
\vskip2mm
% apparently the height/width is set first before angle in epsfig
\centerline{\epsfig{figure=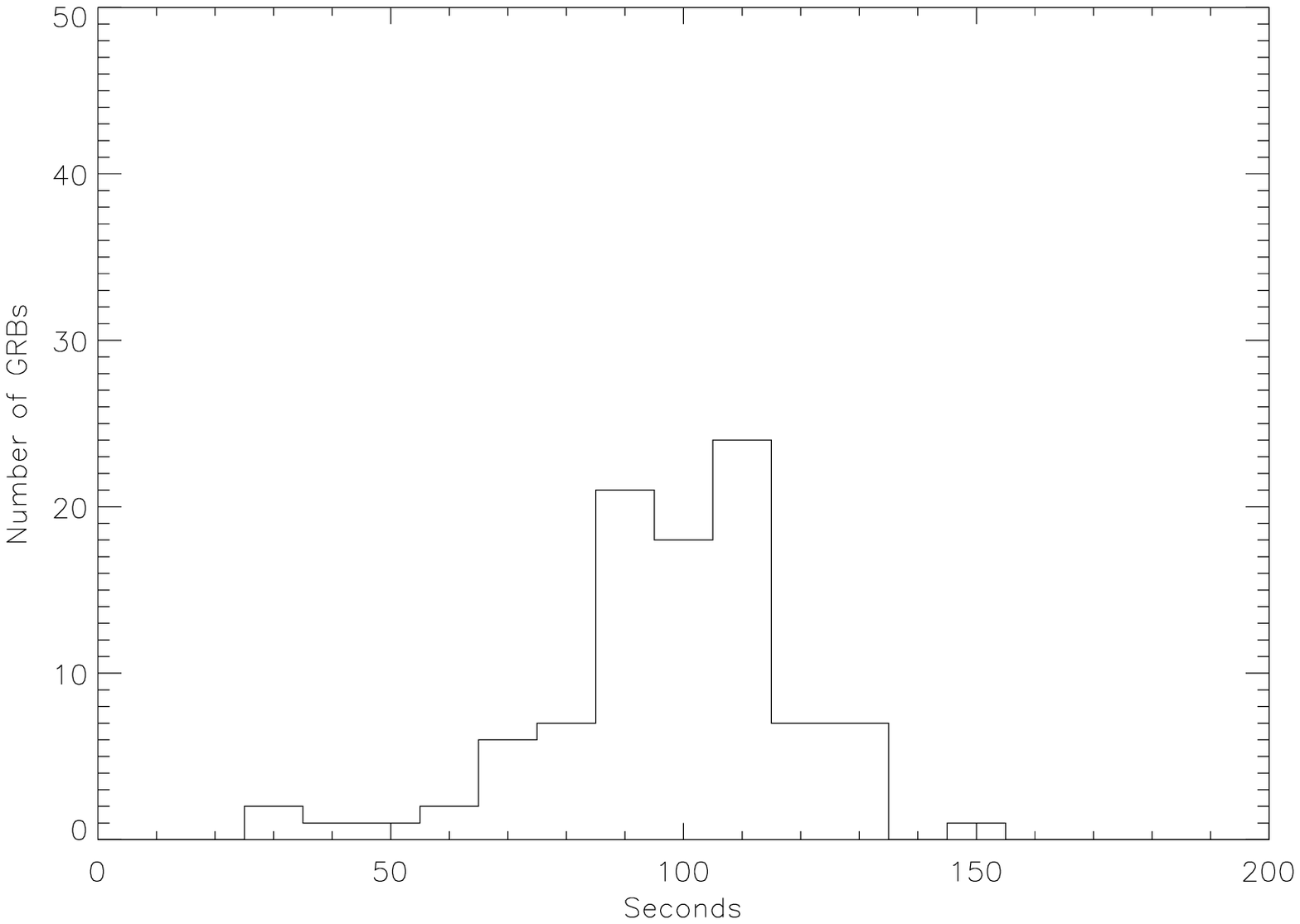,width=90truemm,angle=0,clip=}}
\captionb{3}{The time from the peak of the GRB to starting the x-ray/UV/optical observations. }
\vskip2mm
\end{figure}

\sectionb{3}{DISCUSSION AND CONCLUSIONS}

 Simulations cannot reveal the unexpected problems that might occur on orbit. At best, they can provide
an upper limit of what can be achieved against known situations. For example, 
during its first year of operation. the High Energy Transient Experiment 
(HETE) operated less than 10\% of the expected time and, therefore, saw many fewer GRBs than predicted
by simulations prior to launch. 

 The 1000 burst simulation reported here tells us the accuracy of which we
can locate GRBs with the BAT (50\% within 1.25 arcmin), but on-orbit
misalignments could certainly affect the accuracy. During the early
on-orbit operations, it should be easy to confirm the BAT alignment to the
spacecraft. It will be more difficult to handle movements of the focal
plane and aperture. We triggered on about 300 bursts but that assumes the
satellite is enabled to operate all the time and only the SAA prevents
triggering. In fact, in this simulation we used very mild SAAs and true
SAAs would probably reduce the number of events. We located 148 of those
bursts and were able to slew to 103 of them. The number that we slew to
depends on spacecraft slewing constraints.  There were 16 events that were
too close to the sun to allow a slew. The constraint that we avoid the
earth's limb by 30 degrees prevented slewing to others. No particular
effort was made in the simulation to optimize where the spacecraft was
pointed while waiting for a GRB.

\begin{figure}
\vskip2mm
% apparently the height/width is set first before angle in epsfig
\centerline{\epsfig{figure=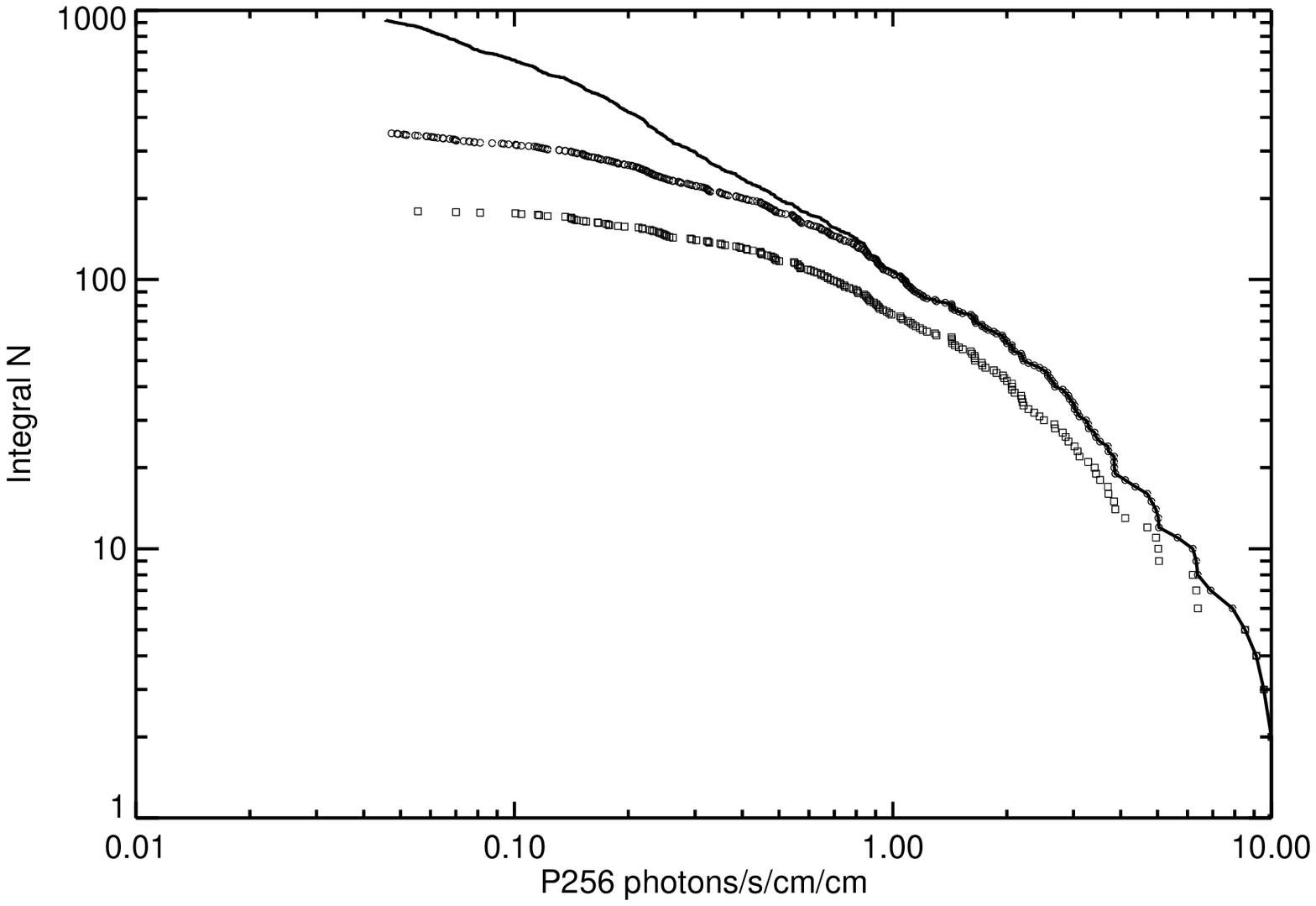,width=90truemm,angle=0,clip=}}
\captionb{4}{Log N Log P based on 1000 Burst simulation that used an improved set of triggers, a likely
set to be flown. Shown are the distribution of simulated bursts, bursts that caused triggers,and bursts that 
were located. There was no simulation of slews. With this set of triggers, we located 184 bursts in about a year assuming that the satellite was constantly in a mode for BAT to locate events. This figure was not in the 
proceedings paper.}
\vskip2mm
\end{figure}

\begin{figure}
\vskip2mm
% apparently the height/width is set first before angle in epsfig
\centerline{\epsfig{figure=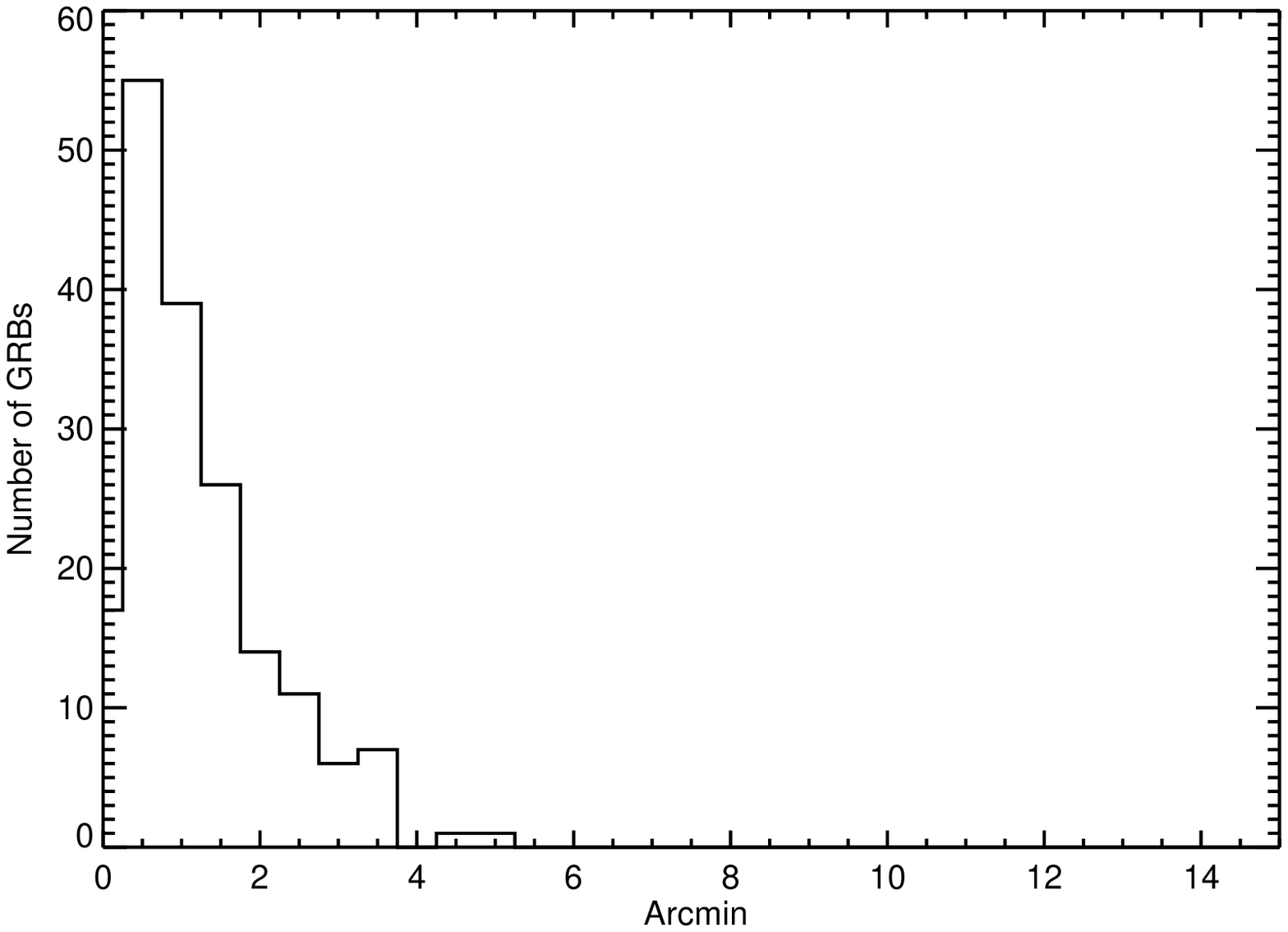,width=90truemm,angle=0,clip=}}
\captionb{5}{Accuracy of on-board locating based on 1000 Burst simulation that used an improved set of triggers, a likely
set to be flown. The accuracy was somewhat improved over that initially reported (i.e., Fig. 2). This figure was not in the 
proceedings paper.}
\vskip2mm
\end{figure}

\begin{figure}
\vskip2mm
% apparently the height/width is set first before angle in epsfig
\centerline{\epsfig{figure=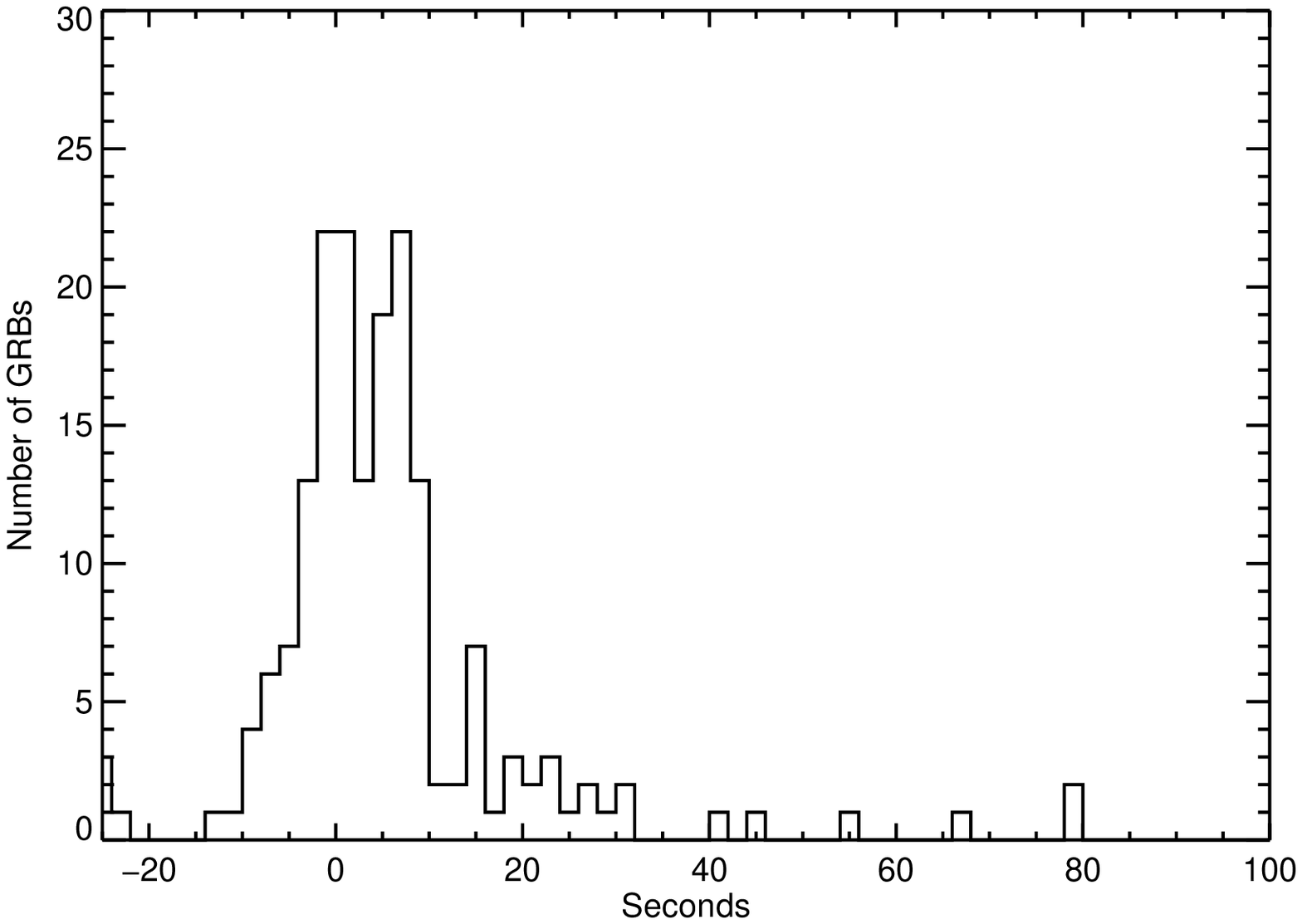,width=90truemm,angle=0,clip=}}
\captionb{6}{The distribution of time between when BAT triggers and the peak of the burst. We triggered on about half of the bursts
before their peaks. In two cases, we trigger
about 80 sec before the peak of the burst.
% Such long bursts will afford the possibility of slewing the XRT and UVOT to observe the bursts while the prompt emission is still active.
This figure was not in the proceedings paper.}
\vskip2mm
\end{figure}

\begin{figure}
\vskip2mm
% apparently the height/width is set first before angle in epsfig
\centerline{\epsfig{figure=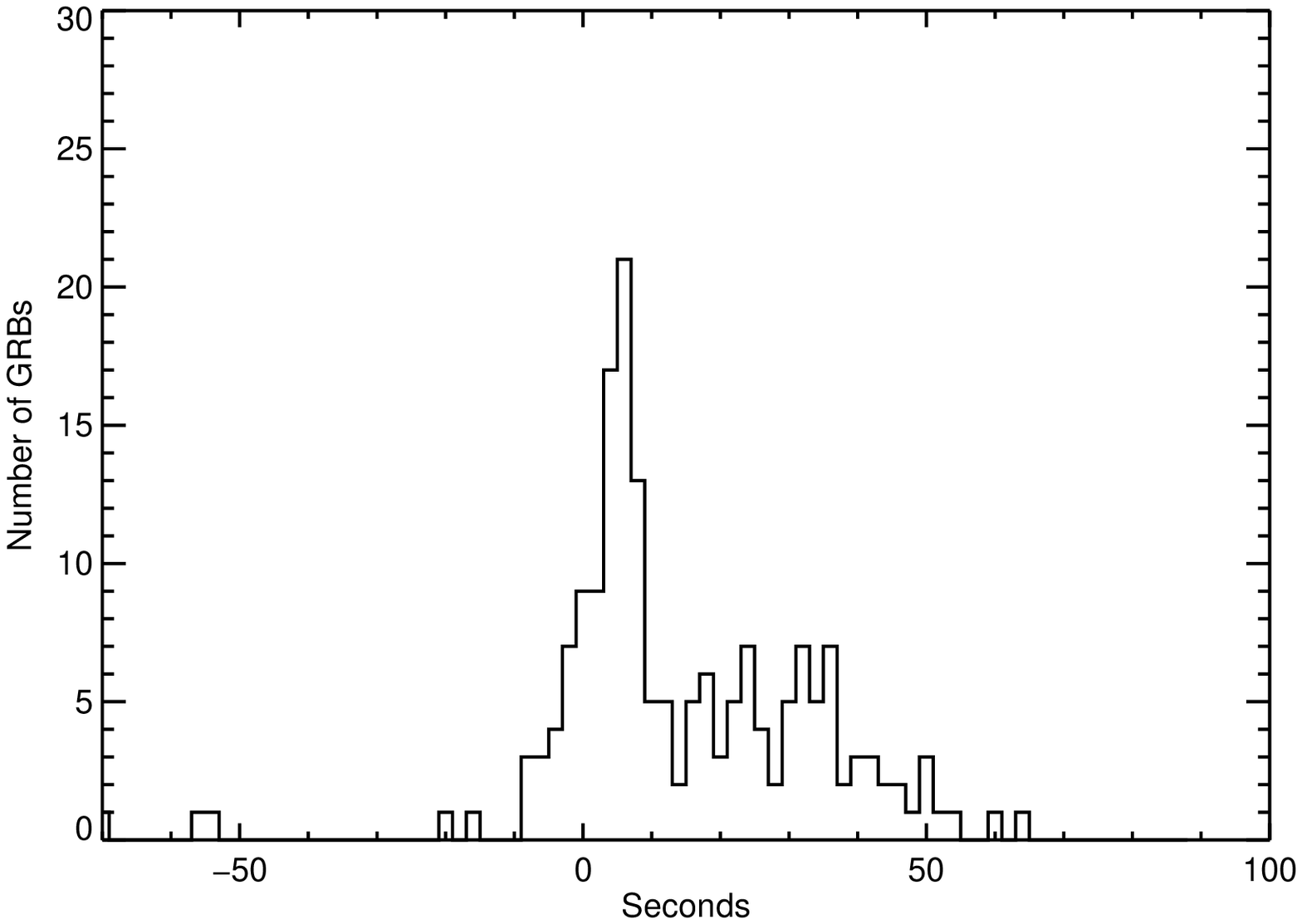,width=90truemm,angle=0,clip=}}
\captionb{7}{The distribution of time between the peak of the burst and when BAT located it.
Although we locate some of the bursts before the peak of the prompt emission, most of the bursts
are located after the peak.
This figure was not in the proceedings paper.}
\vskip2mm
\end{figure}

\begin{figure}
\vskip2mm
% apparently the height/width is set first before angle in epsfig
\centerline{\epsfig{figure=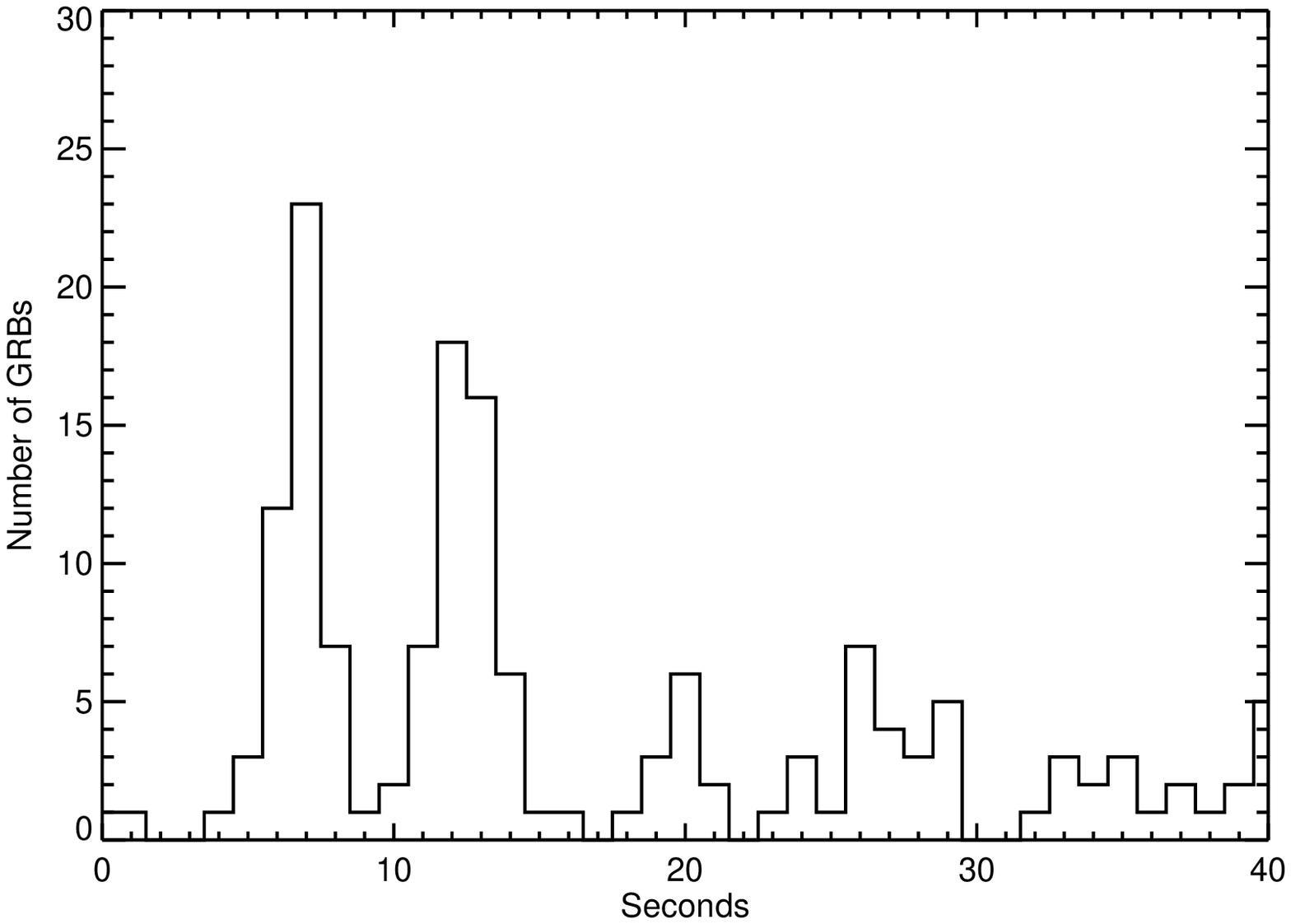,width=90truemm,angle=0,clip=}}
\captionb{8}{Time between the initial BAT trigger and when the BAT location was found. The imaging software was improved
after this proceedings paper was prepared, now the on-board imaging takes about 6 seconds, rather than 12
seconds. The locating software continues to utilize new foregrounds and background combinations
until a new source is located in the net image. In this simulation, a location was found in the first attempt about 1/3 of the
bursts, it took two attempts for another 1/3 of the bursts, and more than two locations for the remainder of the bursts.
Note that we tally the time from the trigger, not the "burst" which has a poorly defined occurrence time.
This figure was not in the 
proceedings paper.}
\vskip2mm
\end{figure}

\begin{figure} \vskip2mm
\centerline{\epsfig{figure=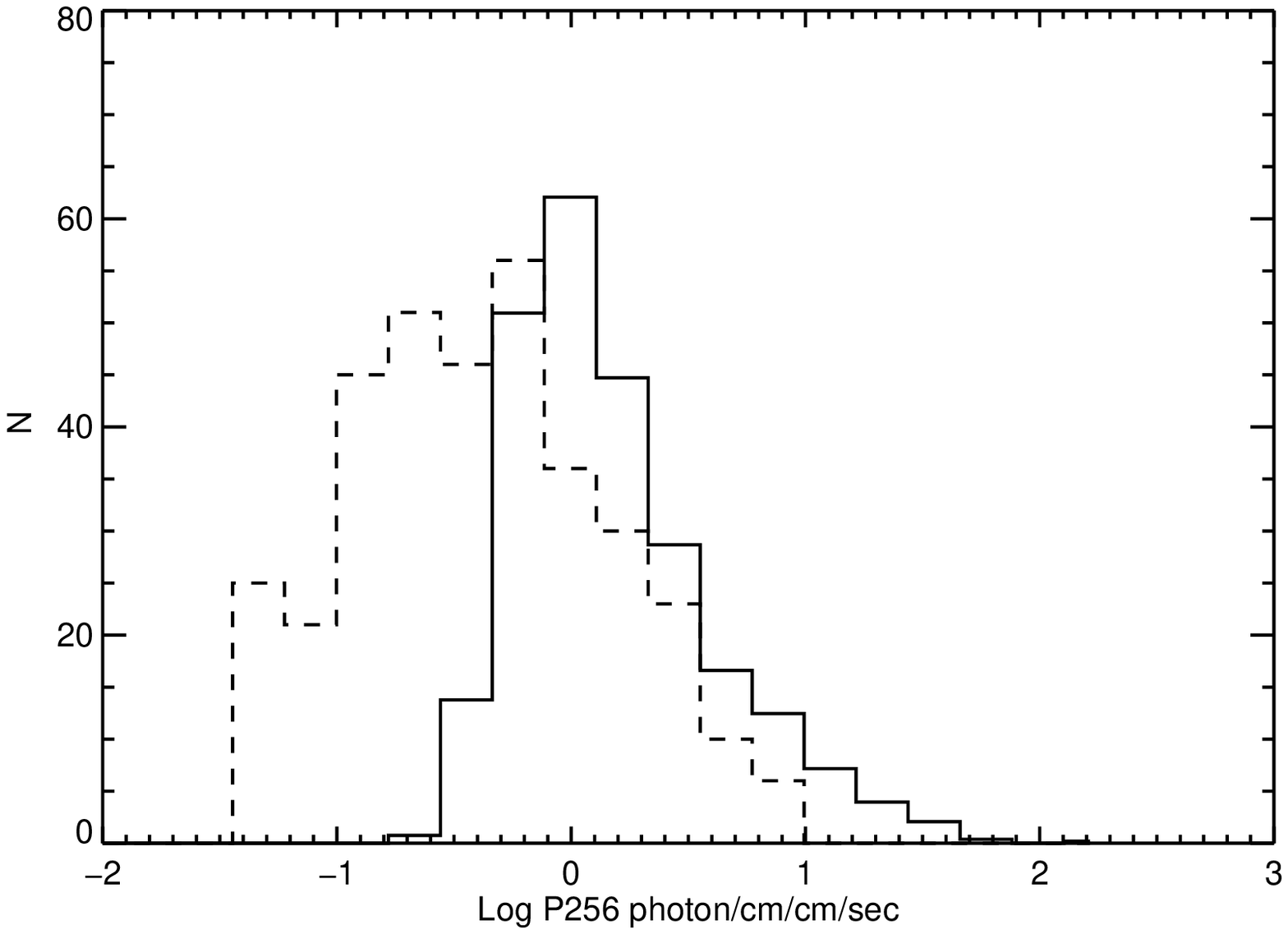,width=90truemm,angle=0,clip=}}
\captionb{9}{Distribution of peak fluxes for detected gamma-ray bursts. 
The solid line is that detected by BATSE per year. The
dashed curve is for the (simulated) events detected (i.e., triggered) by BAT in roughly a year. 
Thus, BAT can detect events about 5 times fainter than BATSE. This figure was not in the
proceedings paper.
} \vskip2mm \end{figure}

\begin{figure} \vskip2mm
\centerline{\epsfig{figure=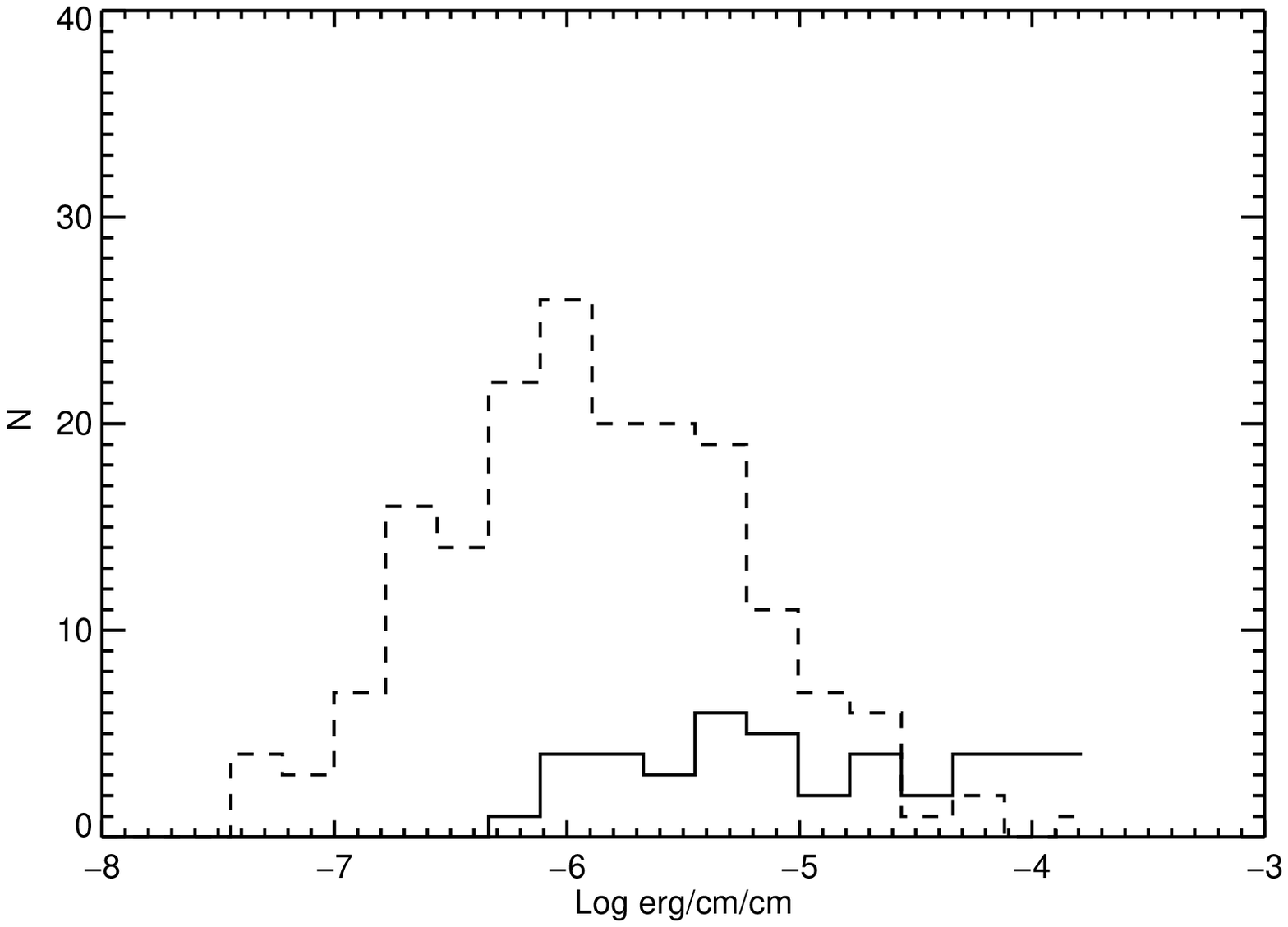,width=90truemm,angle=0,clip=}}
\captionb{10}{Distribution of fluences for located gamma-ray bursts. 
The solid line is for counterparts detected since 1997 by BeppoSax, HETE, the IPN, etc. (data taken from
Friedman and Bloom, in preparation).
The
dashed curve is for the (simulated) events located by BAT per year. 
Thus, BAT can locate events with about 10 times lower fluences than previous gamma-ray burst counterparts.
This figure was not in the
proceedings paper.} \vskip2mm \end{figure}

\goodbreak

\References

\noindent 
Barthelmy,~S. et al.\ 2003 SPIE, 5165-21, in press

\noindent 
Fenimore,~E.~E. et al.\ 2003 in {\smalit Gamma-Ray Burst and Afterglow 
Astronomy}, AIP Conf. 662, eds. G. Ricker \& R. Vanderspek, p.\ 491 

\noindent 
Gehrels,~N. et al.\ 2003, ApJ submitted. 

\noindent
McLean,~K. et al.\ 2004, in {\smalit Gamma-ray Bursts:
30 years of Discovery}, AIP Conf, eds. E.~E.~Fenimore \& M.~Galassi

\noindent
Palmer,~D. 2004, et al.\ 2004, in {\smalit Gamma-ray Bursts:
30 years of Discovery}, AIP Conf, eds. E.~E.~Fenimore \& M.~Galassi

\end{document}